\documentclass[twocolumn,10pt]{article}
\usepackage{graphicx}
\newcommand{\qqbar}{q\bar{q}}
\newcommand{\bbbar}{b\bar{b}}
\newcommand{\ppbar}{p\bar{p}}
\newcommand{\ttbar}{t\bar{t}}
\newcommand{\BR}{\mbox{BR}}
\begin{document}
\title{ 
  STANDARD MODEL HIGGS SEARCHES 
  }
\author{
  Stefan S\"oldner-Rembold\\
  {\em University of Manchester, Oxford Road, Manchester, M13 9PL,
  United Kingdom} \\
  }

%\maketitle

%\baselineskip=11.6pt

  \twocolumn[
    \begin{@twocolumnfalse}
      \maketitle
      \begin{abstract}
The status and perspectives of Standard Model Higgs searches
are discussed.

\bigskip
      \end{abstract}
    \end{@twocolumnfalse}
  ]

%
%\newpage
\baselineskip 14pt

\section{Introduction}
In the Standard Model (SM) the Higgs mechanism is responsible
for breaking  electroweak symmetry, thereby giving mass to the
the $W$ and $Z$ bosons. It predicts the existence of a heavy scalar boson, the
Higgs boson, with a mass that can not be predicted by the SM. 
\begin{figure}[htb] 
\begin{center} 
\includegraphics[width=0.45\textwidth]{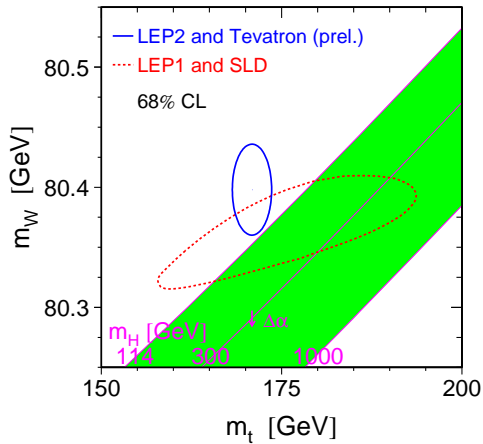} 
\end{center} 
\caption[]{
Dependence of the Higgs
mass on the measured $W$ and top masses.
}
\label{fig-ew1} 
\end{figure} 

\begin{figure}[htb] 
\begin{center} 
\includegraphics[width=0.45\textwidth]{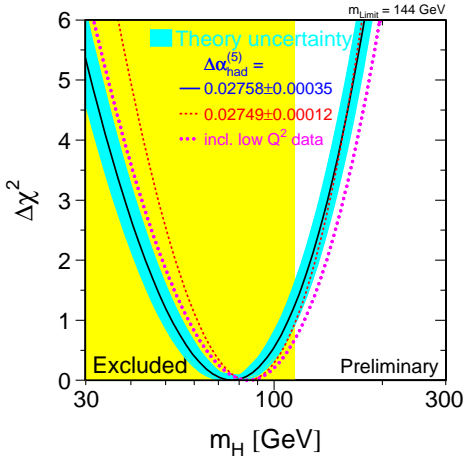} 
\end{center} 
\caption[]{
$\Delta\chi^2$ curve derived from precision electroweak data
as a function of Higgs mass~\protect\cite{bib-lep}.
}
\label{fig-ew2} 
\end{figure}

The LEP, SLD and Tevatron experiments constrain the mass of the
SM Higgs boson indirectly through electroweak precision
measurements. The main contribution to these indirect constraints
from the Tevatron experiments D\O\ and CDF are the measurements of the $W$ and top 
masses~\cite{bib-ew}. The dependence of the Higgs
mass on these measurements is shown in Figure~\ref{fig-ew1} and 
the Higgs mass dependence on the measured 
electroweak precision
parameters in Figure~\ref{fig-ew2}. The SM fit yields a best value of
$m_H=76^{+36}_{-24}$~GeV~\cite{bib-lep}. The upper limit on the Higgs
mass at $95\%$ Confidence Level (CL)\footnote{All limits given
in this paper are at $95\%$ CL} is $m_H<144$~GeV. 

The LEP experiments have searched for Higgs production in
the process $e^+e^-\to ZH$ at $e^+e^-$ centre-of-mass energies $\sqrt{s}$ 
of up to $206.6$~GeV. The maximum kinematically accessible Higgs
mass is therefore $m_H=\sqrt{s}-m_Z=115.4$~GeV. The direct mass limit is
below this values, at $m_H=114.4$~GeV~\cite{bib-lephiggs}, due to the slight
excess observed in the final LEP data.
The ALEPH experiment had reported an excess in this mass region
compatible with the production of a Higgs boson. After combining
the data of all four LEP experiments, the confidence level for
the background hypothesis, $1-CL_b$, is 0.09 while the confidence
level for the signal plus background hypothesis, $CL_{s+b}$, is 
0.15~\cite{bib-lephiggs}.
Taking into account the direct mass limit the upper limit on $m_H$ 
from the electroweak fit increases to $m_H<182$~GeV.

\section{Higgs Searches at the Tevatron}
The Tevatron experiments CDF and D\O\
search for direct Higgs boson production in the mass range
above the LEP limit using $\ppbar$ collisions at $\sqrt{s}=1.96$~TeV.
The relevant processes at these energies are associated Higgs production 
($qq'\to WH$, $q\bar{q}\to ZH$) and gluon fusion ($gg\to H$). 
Typical cross-sections are $\sigma \simeq 0.7-0.15$~pb for gluon fusion
and $\sigma \simeq 0.2-0.02$~pb for associated production at Higgs masses
in the range $115-200$~GeV. 

The Higgs boson predominantly decays into $b$ quarks in the low mass range 
around $115$~GeV. The signal in the $gg\to H$ channel
is therefore overwhelmed by multi-jet background.
The process $gg\to H$ is therefore not a viable
search channel at low masses. The $WH$ and $ZH$ channels with the vector boson
decaying into leptons have much lower cross-sections
but the lepton tag from the decay of the $W\to \ell\nu$ or $Z\to\ell\ell$ and 
selections on missing transverse energy from the neutrino in the
decays $W\to \ell\nu$ or $Z\to\nu\nu$  
help to reduce the background significantly.

At higher masses, around $m_H=160$~GeV, the Higgs boson will predominantly
decay into $WW$ pairs. Leptons from the decays of the $W$ bosons and the missing
transverse energy are used to reject background, making the channel $gg\to
H\to WW$ the most promising search channel in this mass region.
In addition, a 'hybrid' channel, associated production with subsequent Higgs decay into
(virtual) W pairs, $qq'\to WH \to WWW$, also contributes significantly
in the intermediate mass region.

In August 2007, the two Tevatron experiments
have each recorded about 2.8 fb$^{-1}$ of integrated luminosity. Most results
presented here are based on data sets corresponding to  $1-1.9$~fb$^{-1}$,
superseding most of the published results~\cite{bib-oldd0,bib-oldcdf}.
The preliminary results of the two collaborations are
accessible through their web pages~\cite{bib-web}.

\subsection{The Tools}
The main tools employed in Higgs searches at the Tevatron
are lepton identification and - mainly in the low Higgs mass region - 
jet reconstruction and $b$ jet tagging.
The experiments apply $b$ jet tagging algorithms that exploit the long lifetime
of $b$ hadrons. 
These algorithms are applied to each jet, searching for tracks with large transverse 
impact parameters relative to the primary vertex and for 
secondary vertices formed by tracks in the jet.
\begin{figure}[htbp] 
\begin{center} 
\includegraphics[width=0.45\textwidth]{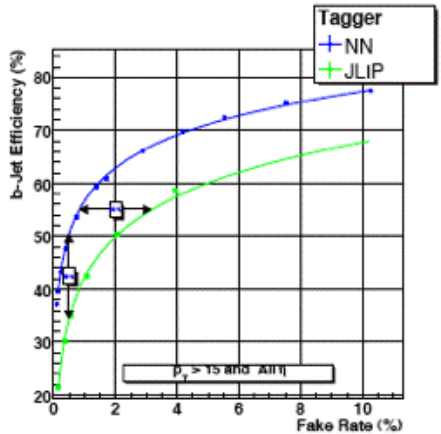} 
\end{center} 
\caption[]{
Dependence of the $b$ tagging efficiency per jet on the
rate of misidentified jets (fake rate) for different
operating points of the NN and the JLIP tagging algorithms
used by the D\O\ experiment.
}
\label{fig-btag} 
\end{figure} 

For advanced $b$ tagging these variables are  used as input to a Neural Network (NN).  
The NN is trained
to separate $b$ quark jets from light flavour jets. By adjusting the minimum requirement
on the NN output variable, a range of increasingly stringent $b$ tagging operating points
is obtained, each with a different signal efficiency and purity. 
The improved performance of a NN tagging algorithm over an algorithm
using jet lifetime probabilities (JLIP) is shown in Figure~\ref{fig-btag}.

Events with neutrinos in the final state are identified
using missing transverse energy. The reconstruction
of all these variables require excellent performance
of all detector components.

\subsection{Signal and Background}

The Higgs signal is simulated with \
PYTHIA~\cite{bib-pythia}. The signal cross-sections are normalised to 
next-to-next-to-leading order (NNLO) calculations~\cite{bib-nnlo} and branching
ratios from HDECAY~\cite{bib-hdecay}.

There are different types of background to the Higgs search.
An important source of background are
multi-jet events (often labeled ``QCD background''). This background
and the instrumental
background due to mis-identified leptons or $b$ jets is simulated
with PYTHIA (only for the CDF $ZH\to\nu\nu\bbbar$ analysis) 
or is taken directly from data, since
it is not very well simulated by Monte Carlo.
Determining this background from data is done 
using control samples with no signal content.

Electroweak background processes such as di-boson production, 
$\ppbar\to VV (V=W,Z)$,
$V$+jets or $\ttbar$ pair production often dominate at the final stages
of the selection; these are simulated using leading order Monte Carlo
programs such as PYTHIA, ALPGEN, HERWIG or COMPHEP.
The normalisation of these processes is obtained either from
data or from from NLO calculations.

\subsection{$ WH \to \ell \nu \bbbar$}

The final state in the channel $ WH \to \ell \nu \bbbar$ 
consists of two $b$ jets from the Higgs boson and
a charged lepton $\ell$ (electron or muon) and a neutrino from the $W$ boson.
The decay mode where the charged lepton is a $\tau$ lepton
decaying into $e,\mu$ is included. The hadronic $\tau$ decay modes
are being studied but are not included yet.

\begin{figure}[htb] 
\begin{center} 
\includegraphics[width=0.4\textwidth]{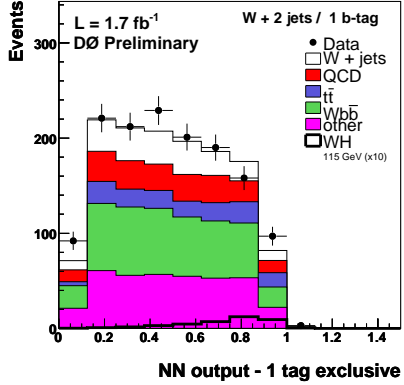} 
\includegraphics[width=0.4\textwidth]{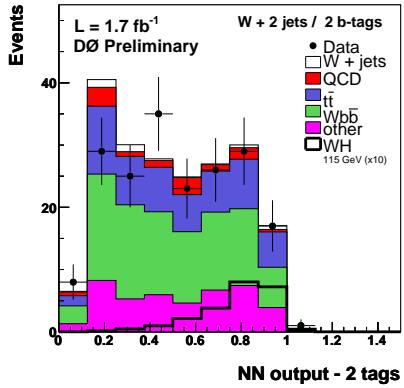} 
\end{center} 
\caption[]{ $WH$ channel:
distribution of the NN output in the two-jet sample for data compared
to the expectation for background and signal: a) exclusive single $b$ tag channel;
b) double $b$ tag channel compared to
a signal expectation for $m_H=115$~GeV.
}
\label{fig-whd0} 
\end{figure} 

D\O\ and CDF therefore select events with one or two tagged $b$ jets with a 
transverse momentum $p_T>20$~GeV, 
one isolated electron or muon with $p_T>15$~GeV (D\O ) or $20$~GeV (CDF)
and  missing transverse energy $E_T^{\rm miss}>20$ GeV. 
The main backgrounds after the selection are $W$ production in association with
two heavy flavour jets and $\ttbar$ production. 

In the CDF analysis a secondary vertex $b$ tag is required for the first jet and
then two independent samples are created by requiring either a secondary vertex
or a jet probability $b$ tag for the second jet. 
D\O\ splits the data sample
into one sample where there is exactly one jet tagged by a tight operating
point of the NN $b$ tagging algorithm and a second sample where two jets are
tagged using a loose operating point. 
Since the samples are further subdivided
by lepton flavour, there are four independent samples in the final analysis.
This improves the sensitivity of the analysis by exploiting the different
signal over background ratios of these selections.

\begin{figure}[htb] 
\begin{center} 
\includegraphics[width=0.4\textwidth]{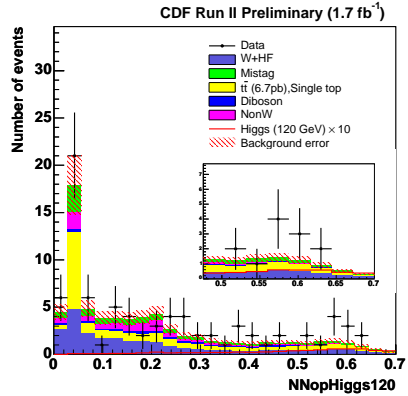} 
\includegraphics[width=0.4\textwidth]{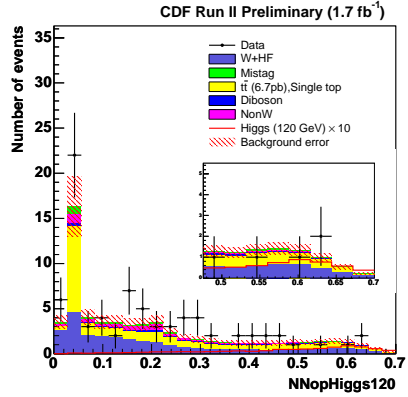}
\end{center} 
\caption[]{$WH$ channel:
kinematic NN output distribution for events with a) one silicon vertex or
jet-probability $b$ tag and b) two silicon vertex $b$ tags compared to
a signal expectation for $m_H=120$~GeV.
}
\label{fig-whcdf} 
\end{figure} 

Kinematic NNs are trained on the selected events to separate a potential Higgs signal
from the background. The NN output for the D\O\ data is shown in Figure~\ref{fig-whd0}
for the two-jet sample for events with one exclusive $b$ tag and with
two $b$ tags. The signal expectation for a Higgs with $m_H=115$~GeV 
is peaked at large NN output values. Similar distributions are shown for
the CDF data in Figure~\ref{fig-whcdf}.

Both experiments use about 1.7 fb$^{-1}$ of data for this channel.
To compare the sensitivity of the different channels directly, all limits are
expressed as a ratio with respect to the SM cross-section, in this channel
$\sigma_{SM}=\sigma(\ppbar\to WH)\BR(H\to\bbbar)$.
The $WH$ search yields a median expected (observed) upper limit on the $WH$ production cross-section of
$\sigma_{95}/\sigma_{SM}=9.95 (10.1)$ for CDF and $9.05 (11.1)$ for D\O\ at a Higgs mass
of $m_H=115$~GeV.

\subsection{$ZH \to \nu\nu \bbbar$}
\begin{figure}[htbp] 
\begin{center} 
\includegraphics[width=0.45\textwidth]{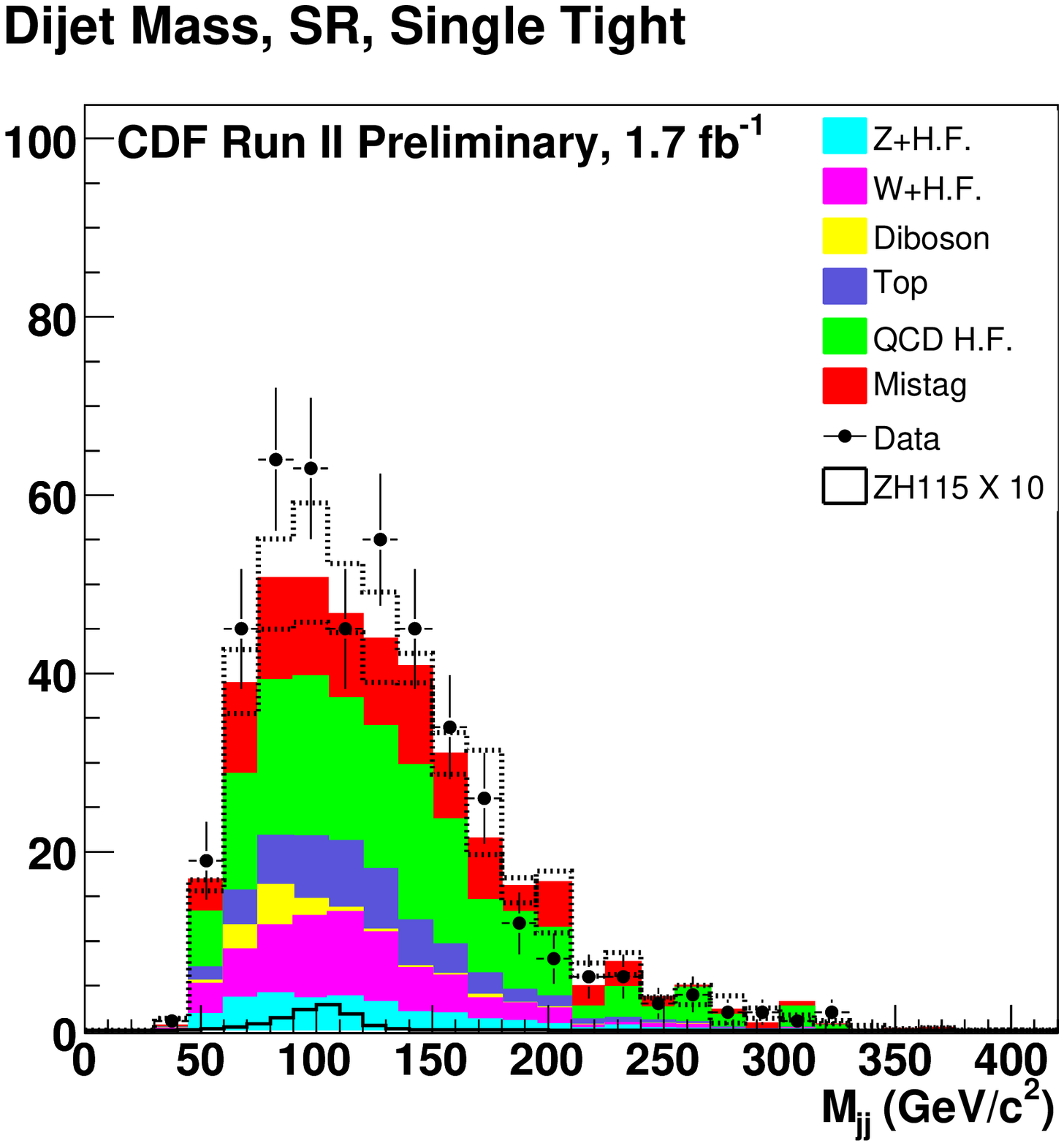}
\includegraphics[width=0.45\textwidth]{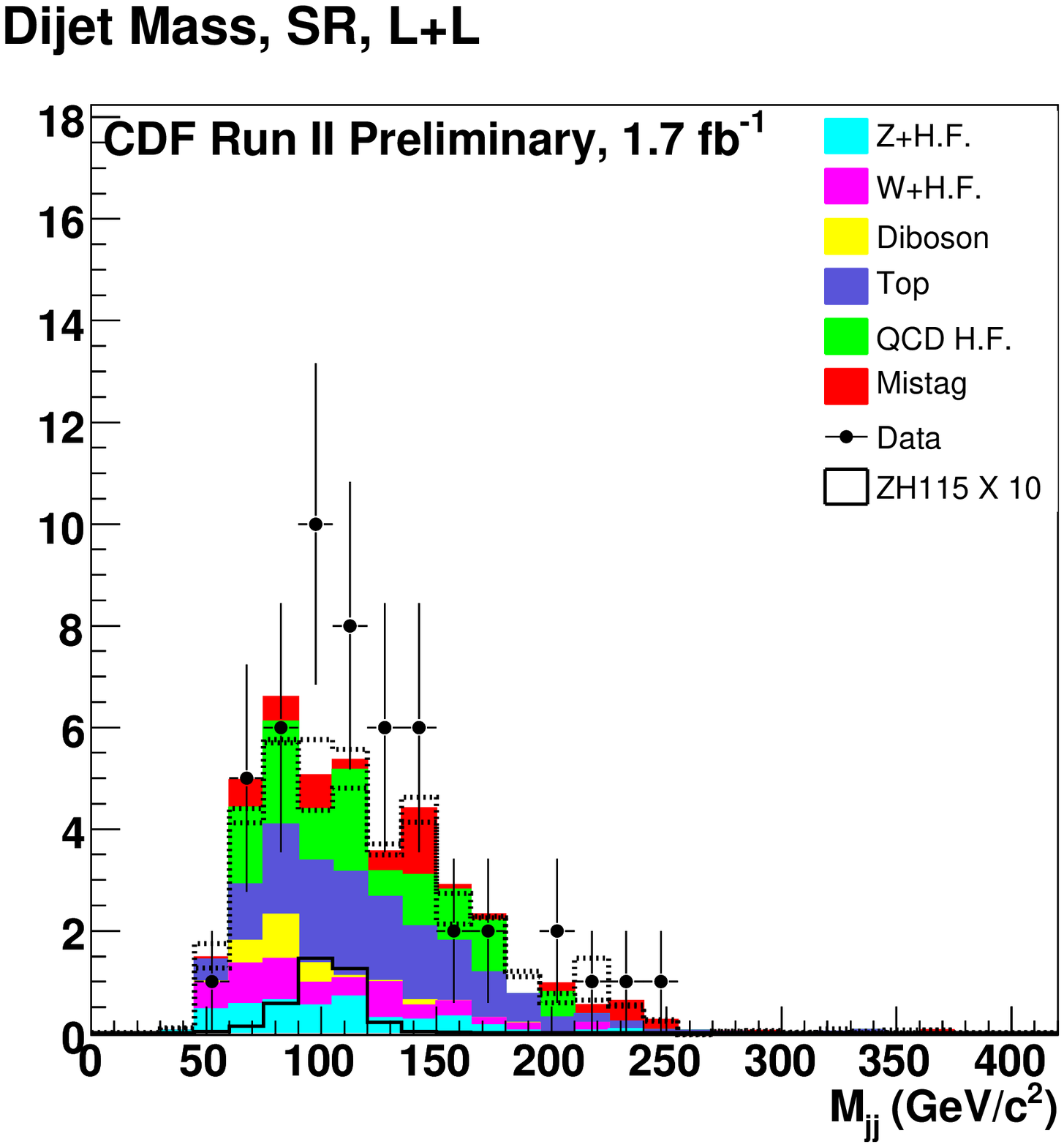}
\end{center} 
\caption[]{$ZH\to\nu\nu\bbbar$ channel:
di-jet mass distribution for data compared to
the background expectation with a) a single tight operating point $b$ tag and
b) two loose $b$ tags with a signal distribution (x10) for $m_H=115$~GeV.
}
\label{fig-metcdf} 
\end{figure} 
This channel has
very good sensitivity because of the large branching ratios
for $Z\to\nu\nu$ and $H\to\bbbar$ decays. Since the two b-jets
are boosted in the transverse direction, the signature for
the final state are acoplanar di-jets, in contrast to 
most background di-jet events which are expected to be back-to-back in
the transverse plane, and large missing transverse
energy. 
Main background sources are $W$ or $Z$ production in association
with heavy flavour jets, multi-jet events and $\ttbar$ pairs. 

The basic selection requires at least one (CDF) or two jets (D\O ) with 
a $b$ tag, large missing transverse energy
(D\O : $E_T^{\rm miss}>50$~GeV, CDF: $E_T^{\rm miss}>70$~GeV),
and a veto on any isolated muon or electron in the event.
\begin{figure}[htbp] 
\begin{center} 
\includegraphics[width=0.45\textwidth]{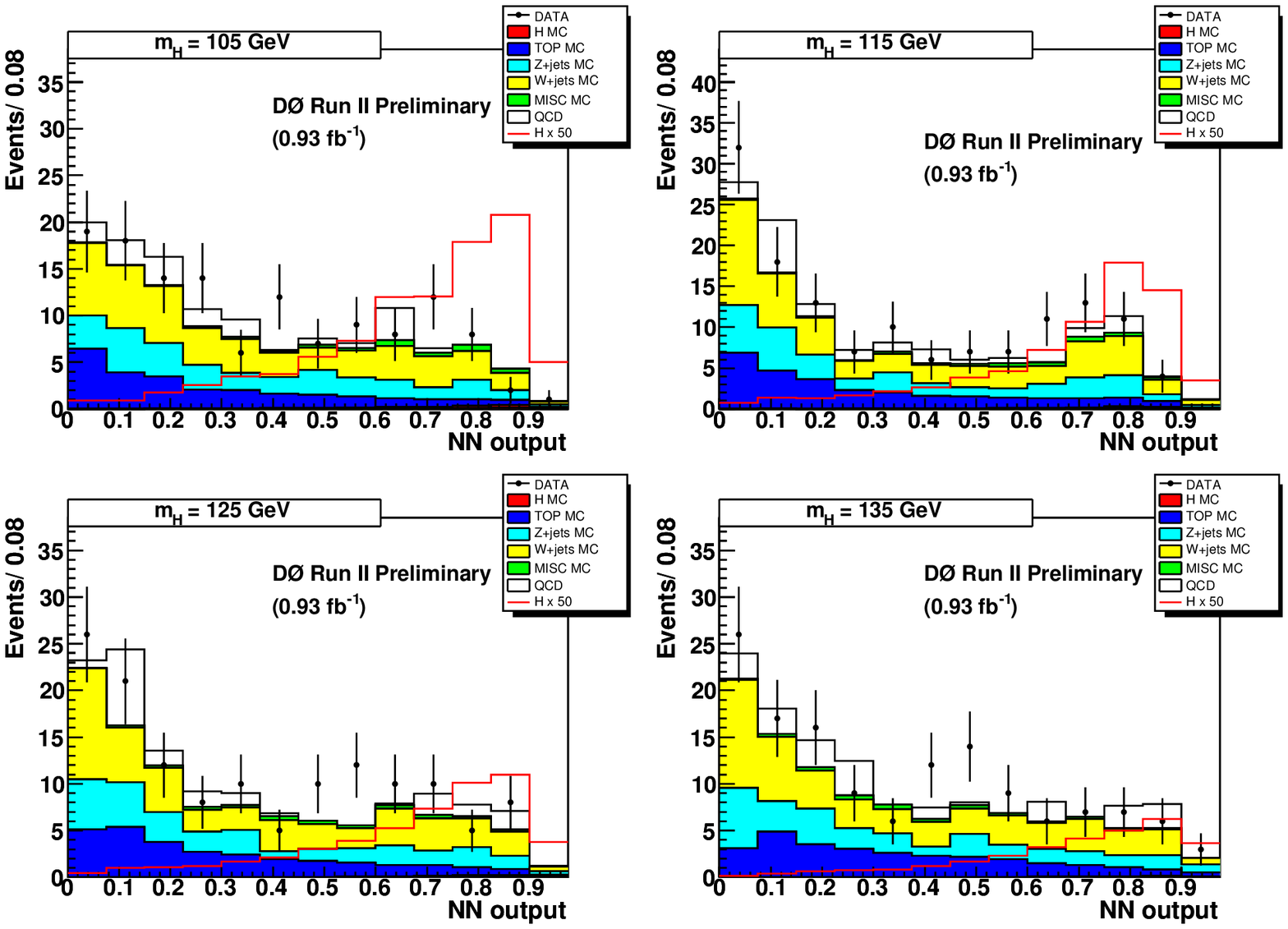}
\end{center} 
\caption[]{$ZH\to\nu\nu\bbbar$ channel:
Kinematic NN output distribution for data compared to the background
and signal expectations for a range of Higgs masses. 
}
\label{fig-metd0} 
\end{figure} 

In the CDF analysis, the final sample is divided into two samples,
one sample with exactly one tight secondary vertex $b$ tag and
a second sample with two loose secondary vertex $b$ tags. The limit
setting is done using the di-jet invariant mass distribution shown
in Figure~\ref{fig-metcdf}.
In the case of D\O ,  events with two NN $b$ tags are used to
train a kinematic NN for identifying signal events.
Asymmetric operating points, one loose and one tight,
are chosen for the two $b$ tags. The output distributions
of the NN, retrained for every Higgs mass,
is shown in Figure~\ref{fig-metd0}.

To increase the sensitivity of this analysis, $WH$ signal events
where the charged lepton has not been identified are also included
in the signal definition. 
This search yields a median expected (observed) upper limit on the 
$VH (V=W,Z)$  production cross-section of
$\sigma_{95}/\sigma_{SM}=9.7 (19.7)$ for CDF and $12.3 (13.1)$ for D\O\ at a Higgs mass
of $m_H=115$~GeV. The CDF data set corresponds to $1.7$~fb$^{-1}$
and the D\O\ data to $1$~fb$^{-1}$
The discrepancy between the median expected and observed limit
for CDF is about $\simeq 2\sigma$. 

\subsection{$ZH \to \ell\ell \bbbar$}

In this channel the $Z$ boson is reconstructed by its decay into 
two high-$p_T$ isolated muons or electrons. The reconstructed $Z$ 
and two b-tagged
jets are then used to select the Higgs signal. 
The invariant mass of the two leptons is required to be in the
$Z$ mass range $70<m_Z<110$~GeV (D\O ) or $76<m_Z<106$~GeV (CDF).
Both experiment require two jets with either one tight $b$ tag or two loose $b$ tags.

\begin{figure}[htb] 
\begin{center} 
\includegraphics[width=0.44\textwidth]{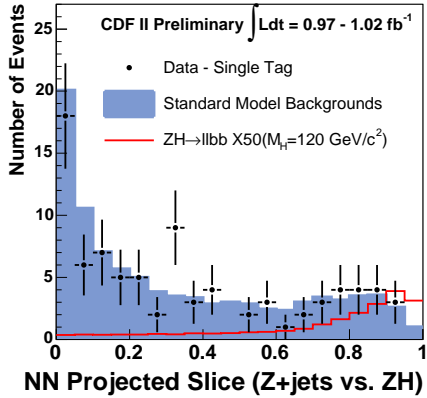}
\includegraphics[width=0.44\textwidth]{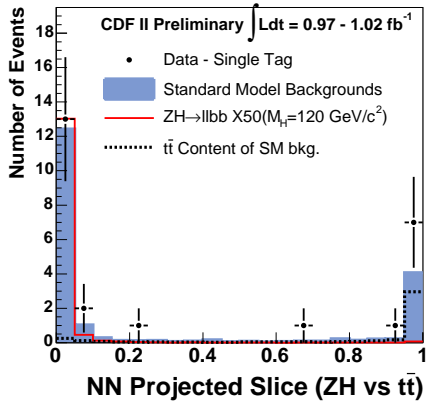}
\end{center} 
\caption[]{$ZH\to\ell\ell\bbbar$ channel:
a) NN output projected on the $y$-axis for $x_{NN}>0.75$ and 
b) NN output projected on the $x$-axis for $y_{NN}<0.2$ for data, background
and signal expectations using $M_H=120$~GeV.
}
\label{fig-llcdf} 
\end{figure} 

The main background sources
are $Z$ production in association with heavy jets and $\ttbar$ production.
$ZZ$ production is an irreducible background apart from the mass
discriminant.
CDF trains two separate NNs to reject these two background components.
Slices of the output of these NNs, 
projected on the two axes, is shown in Figure~\ref{fig-llcdf}.
The di-jet mass resolution is improved by training a different NN
using $E_{T}^{miss}$ and the kinematics of both jets. The data set corresponds
to an integrated luminosity of 1~fb$^{-1}$. 
The D\O\ analysis is performed with $1.1$~fb$^{-1}$ of data using a kinematic
NN and two $b$ tag samples with one tight $b$ tags and two loose $b$ tags.

These searches yield a median expected (observed) upper limit on the $ZH$ production 
cross-section of
$\sigma_{95}/\sigma_{SM}=16 (16)$ for CDF and $20.4 (17.8)$ for D\O\ at a Higgs mass
of $m_H=115$~GeV. Even though the limits are less stringent than
for the $ZH\to\nu\nu\bbbar$ channel, it still provides important input
to increase the overall sensitivity of the analysis.

\begin{figure}[htbp] 
\begin{center} 
\includegraphics[width=0.45\textwidth]{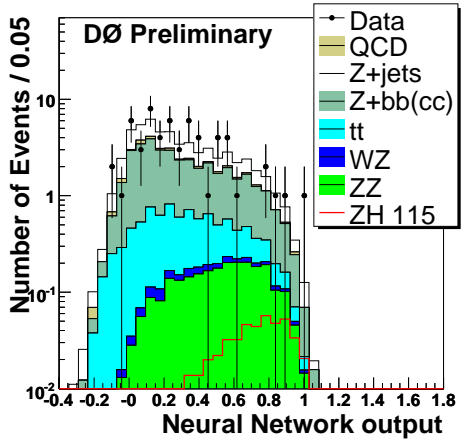}
\end{center} 
\caption[]{$ZH\to\ell\ell\bbbar$ channel:
NN output for the event sample with two loose $b$ tags for data, background
and signal expectations using $M_H=115$~GeV.
}
\label{fig-lld0} 
\end{figure} 

\subsection{$W \to WW  \to \ell\nu\ell\nu$}

\begin{figure}[htbp] 
\begin{center} 
\includegraphics[width=0.45\textwidth]{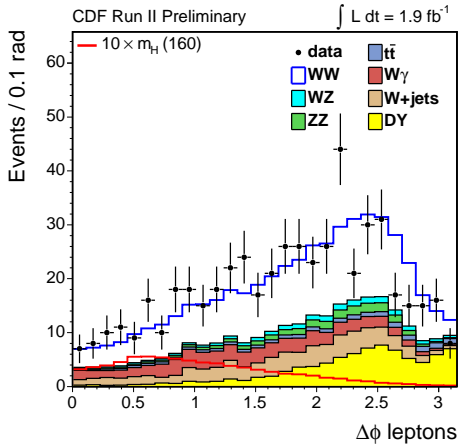}
\end{center} 
\caption[]{$WW$ channel:
azimuthal angle between the two leptons in the $H \to WW$ search. Due to spin
correlations, the signal is at low $\Delta\phi_{\ell\ell}$,
whereas the background is at high $\Delta\phi_{\ell\ell}$.
}
\label{fig-dphi} 
\end{figure}

The dominant decay mode for higher Higgs masses is $H\to WW^{(*)}$.
Leptonic decays of the W bosons are therefore used to suppress QCD background.
The signature of the $gg\to H \to WW^{(*)}$ channel is two high-$p_T$
isolated leptons with small azimuthal separation, $\Delta\phi_{\ell\ell}$,
due to the spin-correlation between the final-state leptons in
the decay of the spin-0 Higgs boson.
In contrast, the lepton pairs from background events, mainly $WW$ events, are predominantly
back-to-back in $\phi_{\ell\ell}$. This is shown in Figure~\ref{fig-dphi} for
a preselected CDF data sample. 
An additional selection requires $E_T^{\rm miss}>25$~GeV to account
for the neutrinos in the final state.

\begin{figure}[htbp] 
\begin{center} 
\includegraphics[width=0.45\textwidth]{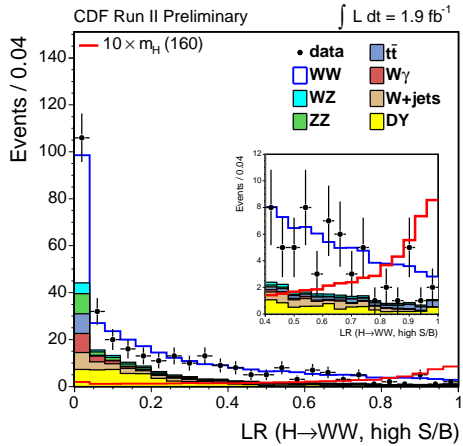}
\end{center} 
\caption[]{$WW$ channel:
Likelihood ratio for calculated for background and a  $H\to WW$ 
signal with $m_H=160$~GeV.
}
\label{LLRWW} 
\end{figure}

The CDF analysis uses an event-by-event probability density 
 \begin{eqnarray*}
  \lefteqn{P_m(\vec{x}_{obs})= }\\
& & \frac{1}{\langle \sigma_m\rangle}\int d^n\sigma_m^{LO}(y)\epsilon(y)G(\vec{x}_{obs},y),\\
  \end{eqnarray*}
where $y$ represents the true lepton kinematics, $\vec{x}_{obs}$ the observed kinematics,
$\epsilon$ the lepton efficiencies and $G$ the detector resolution function. The
probabilities are derived using the leading order cross-sections $\sigma_m^{LO}$
for the signal or background processes $m=s,b$. From the probabilities the 
likelihood ratios
$$LR(m_H)=\frac{P_{s}(m_H)}{P_{s}(m_H)+\sum_{b}f_{b}P_{b}}$$
are calculated as a function of Higgs mass. These distributions are used
for limit setting, an example is shown in Figure~\ref{LLRWW}.

The observed CDF 
limit on the cross-section ratio $\sigma_{95}/\sigma_{SM}$ 
equals 2, while 3.1 is expected for this data set corresponding
to a luminosity of 1.9~fb$^{-1}$.
D\O\ has recently updated their result using 1.7~fb$^{-1}$. The 
resulting cross section ratio limits are 2.4 observed and
2.8 expected at a Higgs mass of 160~GeV~\footnote{
It should be noted that these proceedings and the combination contain
some additional updates which were not available at the time of the conference.
For reasons of consistency and simplicity the status of the data used
for the December 2007 combination is presented throughout these 
proceedings.}

\subsection{$WH \to WWW^{*}  \to \ell \nu \ell^{'} \nu\qqbar$}
In the process $WH \to WWW^{*}  \to \ell \nu \ell^{'}  \nu\qqbar$ the 
Higgs boson is produced in association with a $W$ boson and subsequently decays
into $WW$. This process is important in the intermediate mass range.
The signature is at least two isolated leptons from the $W$
decays with $p_T>15$~GeV and identical charge. The associated $W$
and one of the two $W$ bosons from the Higgs decay should have
the same charge.
For the final signal selection a two-dimensional
likelihood is used, based on the invariant mass of the two leptons,
the missing transverse energy and their azimuthal angular correlations.
The two-dimensional likelihood separates the signal from physics background
and instrumental background, respectively.

This same-sign charge requirement is very powerful in rejecting
background from $Z$ production. The remaining background is either
due to di-boson production or due to charge mis-measurements.
The rate of charge mis-measurements for muons is determined by comparing
the independent charge measurements in the solenoid and in the toroid
of the D\O\ detector. For electrons the charge mis-measurement rate
is determined by comparing the charge measurement from the solenoid
with the azimuthal offset between the track and the calorimeter cluster
associated to the electron.

\begin{figure}[htb] 
\begin{center} 
\includegraphics[width=0.41\textwidth]{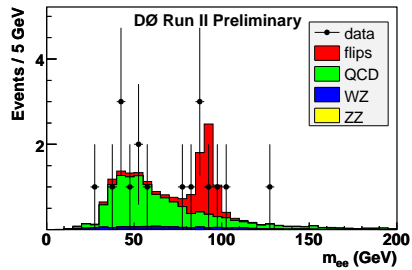}
\includegraphics[width=0.41\textwidth]{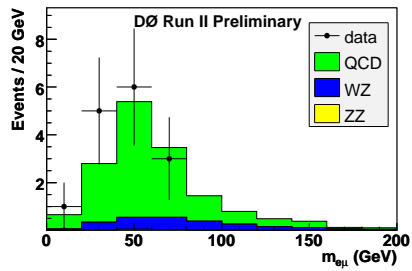}
\includegraphics[width=0.41\textwidth]{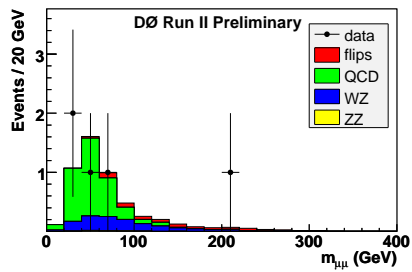}
\end{center} 
\caption[]{$WWW$ channel:
Invariant mass of the two leptons after the final selection
for a) $ee$, b) $e\mu$ and c) $\mu\mu$ events. The background
from charge misidentification (``flips''), the multi-jet background 
(``QCD'') and the di-boson background are shown separately.}
\label{WWW} 
\end{figure} 

The upper limit on the cross-section $\sigma(\ppbar\to W) \BR(H\to WW^*)$
is between 3.2~pb and 2.8~pb for Higgs masses from 115~GeV to 175~GeV.
The expected cross-section ratio in the mass range 140~GeV to 180~GeV
is $\sigma_{95}/\sigma_{SM}\simeq 20$, i.e. this channel makes
a significant contribution at the limit in this mass range. 
A similar analysis was performed by CDF with an integrated luminosity
of 194~pb$^{-1}$~\cite{bib-web}.

\section{Combined Tevatron Limit}
The data of both experiments have been combined using the full
set of analyses with luminosities in the range 
1.0-1.9~fb$^{-1}$~\cite{bib-comb}. 
To gain confidence that the final result does not depend on the details
of the statistical method applied, several types of combination 
were performed,
using both Modified Frequentist (sometimes called the
LEP $CL_s$ method) and Bayesian approaches. The results
agree within about $10\%$. Both methods use Poisson likelihoods and
rely on distributions of the final
discriminants, e.g. NN output or di-jet mass distributions, not only
on event counting.

Systematic
uncertainties enter as uncertainties on the expected number of signal
and background events, as well as on the shape of the discriminant
distributions. The correlations of systematic uncertainties between channels,
different background sources, background and signal and between experiments are taken into account. 
The main sources of systematic uncertainties are, depending on channel,
the luminosity and normalisation, the estimates of the multi-jet backgrounds,
the input cross-sections used for the MC generated background sources,
the higher order corrections ($K$ factors) needed to describe heavy flavour
jet production, the jet energy scale, $b$ tagging and lepton identification.

\begin{figure}[htbp] 
\begin{center} 
\includegraphics[width=0.5\textwidth]{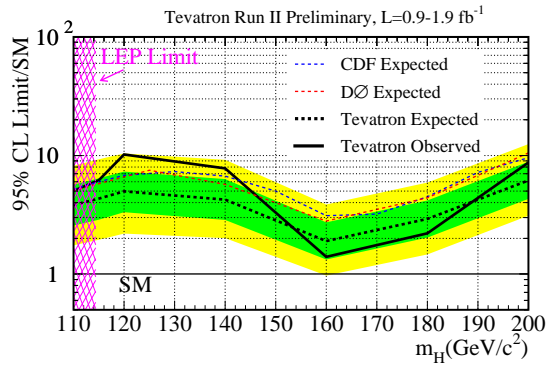} 
\end{center} 
\caption[]{
Expected and observed $95\%$ CL cross-section ratios for the combined
CDF and D\O\ analyses. The median expected $95\%$ CL ratio for both experiments
are also shown (status December 2007).
}
\label{fig-comb} 
\end{figure}

The $95\%$ Confidence Level (CL) 
upper limits on the cross-section 
ratio are shown in Figure~\ref{fig-comb}~\cite{bib-comb}.
The median expected (observed) limits are a factor of 4.3 (6.2) higher than the 
SM cross-sections for a Higgs mass of 115~GeV and a factor 1.9 (1.4) for
a Higgs mass of $160$~GeV. The green and yellow bands show the one and two
sigma bands for background fluctuations. 

The limits are closest to the SM line at a ratio of one for a Higgs mass
of $160$~GeV and it is to be expected that the combined Tevatron experiments will
be able to soon reach the sensitivity needed to make a statement
about the SM Higgs boson in this mass region. 

\section{Higgs Searches at the LHC}
If the SM Higgs boson exists, it will be observed by the LHC experiments
ATLAS~\cite{bib-ATLAS} and CMS~\cite{bib-CMS}. 
Enormous work has gone into optimising Higgs searches
at the LHC and I will only give a very short account here which can
not do justice to this work~\cite{bib-jakobs}.

\begin{figure}[htbp] 
\begin{center} 
\includegraphics[width=0.45\textwidth]{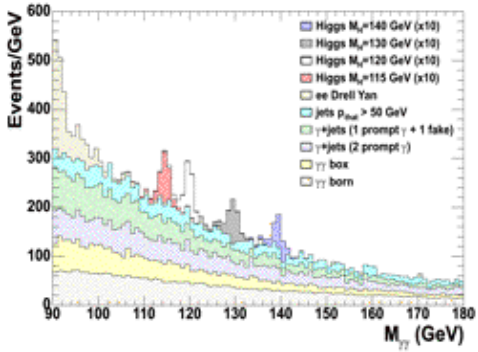} 
\end{center} 
\caption[]{
Di-photon invariant mass spectrum  observed in CMS. The events
are normalised to an integrated luminosity of 1~fb${-1}$ and the
Higgs signals shown for different masses are scaled by a factor ten.
}
\label{fig-cmsgg} 
\end{figure}

The total Higgs production cross-section at the LHC at low masses is about
a factor 50 higher than at the Tevatron. The total cross-section is still
dominated by the process $gg\to H$, but the second most important process
for intermediate and higher Higgs masses is Weak Vector Boson Fusion, $qq\to H qq$.
At small Higgs masses one of the most promising Higgs search channels is
through the decay $H \to \gamma\gamma$. The branching ratio $BR(H\to\gamma\gamma)$
is small, about $0.002$, but in comparison to $\bbbar$ final states it provides
a clear signature that can be distinguished from the large multi-jet background
expected at the LHC. Both ATLAS and CMS have studied this channel which
requires very good photon identification, photon/jet separation and very
good energy resolution. Main backgrounds are di-photon production, $\ppbar\to\gamma\gamma$
and single photon production, $\ppbar\to\gamma+\mbox{jet}$. 
An example for a di-photon mass spectrum is shown in 
Figure~\ref{fig-cmsgg}~\cite{bib-CMS}.

\begin{figure}[htbp] 
\begin{center} 
\includegraphics[width=0.45\textwidth]{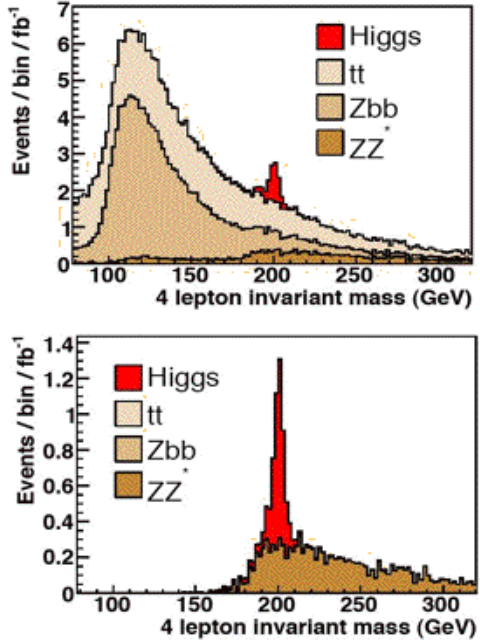} 
\end{center} 
\caption[]{
Four-lepton invariant mass before and after background rejection. 
}
\label{fig-cms4l} 
\end{figure}

The 'golden channel' for higher Higgs masses is the decay $H\to ZZ \to 4$~leptons, where
the leptons are either electrons or muons. In Figure~\ref{fig-cms4l} a
simulation of the signal and background is shown. The irreducible background
is $pp\to ZZ^*$.

\begin{figure}[htb] 
\begin{center} 
\includegraphics[width=0.41\textwidth]{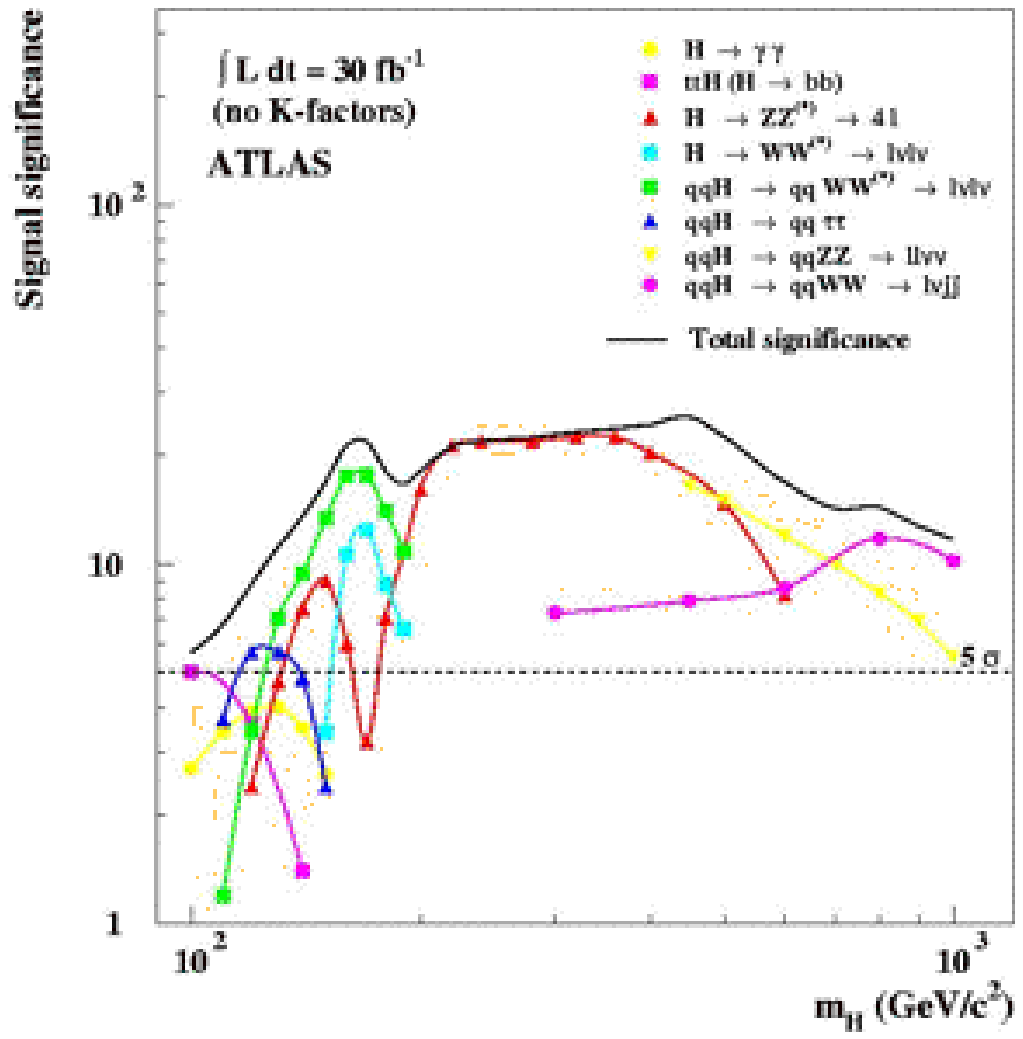} 
\includegraphics[width=0.41\textwidth]{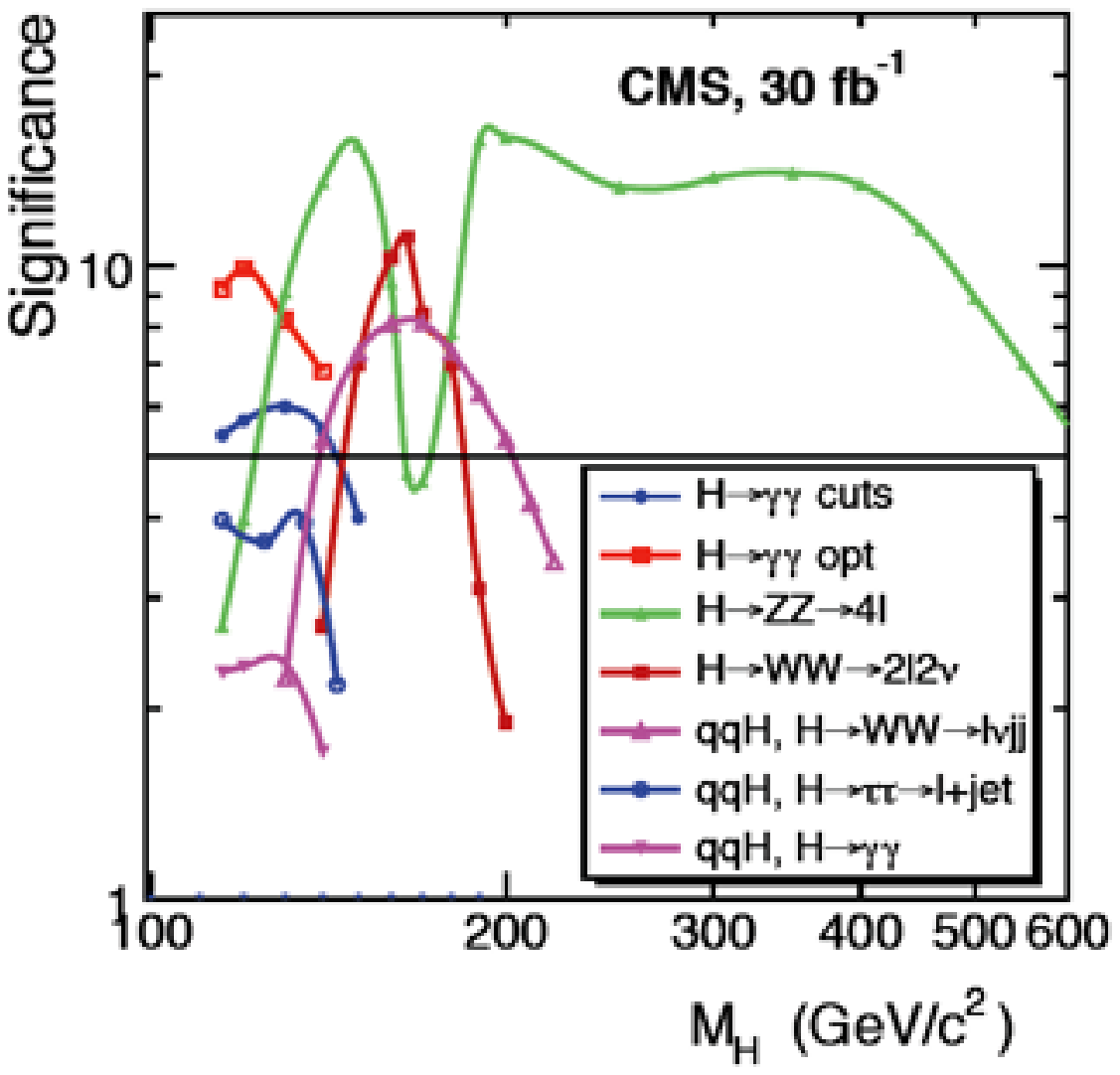} 
\end{center} 
\caption[]{
Expected significance for a SM Higgs discovery for a) ATLAS and b) CMS
as a function of Higgs mass.

}
\label{fig-lhc} 
\end{figure}

The discovery potential (significance) as a function of Higgs
mass for both experiments is shown in Figure~\ref{fig-lhc}, using
all the full range of Higgs production and decay mechanisms for an integrated
luminosity of $30$~fb$^{-1}$. 
The ATLAS study shown here used LO simulation and
has not yet been performed with $k$ factors, whereas the
CMS study includes NLO corrections for signal and background. 
Both experiments can discover the SM Higgs boson with more than five
sigma significance over the full mass range $115<M_H<1000$~GeV.

\section{Summary and Perspectives}
The CDF and D\O\ experiments at Fermilab's Tevatron have searched
for the SM Higgs boson in a variety of channels, using data 
corresponding to integrated luminosities between 0.9~fb$^{-1}$
 and 1.9~fb$^{-1}$.
The observed cross-section limits are about a factor 6.2 above the SM
expectation for a Higgs boson mass of 115 GeV and a factor 1.4 for
a mass of 160~GeV.
 
The Tevatron will continue
to take data in 2009 and possibly in 2010. The total integrated luminosity
is expected to be about $6-7$~fb$^{-1}$ at the end of 2009.
%Recent extrapolations show that with these improvements the combined
%Tevatron experiments are likely to be able to exclude the SM Higgs
%boson in most of the mass range $M_H<200$~GeV or even observe
%three sigma events if the Higgs mass lies in this range.

A long list of
improvements - in addition to adding more data - are expected to
lead to a significant increase in the sensitivity. The most
important improvements are
better $b$ tagging using Neural Net algorithms, improvements of the 
di-jet mass resolution to separate the Higgs signal from the background,
the inclusion of new channels such as $WH\to\tau(\to \mbox{hadrons})\nu
\bbbar$, better lepton identification, the use 
of advanced analysis techniques and improved treatment of systematics. 
These improvements in combination  with increased statistics due to 
more recorded and analysed data will give the Tevatron experiments
a good chance to exclude a SM Higgs boson at $95\%$ CL 
in the low mass region above the LEP limit of 114.4~GeV up to about 200~GeV, or,
if the SM Higgs boson mass lies in this range, to make a $3\sigma$ observation.
The LHC experiments ATLAS and CMS will be be able to discover
the SM Higgs bosons in the full mass range $115<m_H<1000$~GeV
with more than five sigma significance in the first few years
of running.

\section*{Acknowledgements}
The author would like to thank the organisers for making this 
a very enjoyable conference and the Royal Society for 
the conference grant.

\end{document}